# Cointegration and ARDL specification between the Dubai crude oil and the US natural gas market

Stavros Stavroyiannis

Department of Accounting and Finance, School of Management, University of the Peloponnese, GR 241-00, Greece

Email: computmath@gmail.com

**Abstract**

This paper examines the relationship between the price of the Dubai crude oil and the price of the US natural gas using an updated monthly dataset from 1992 to 2018, incorporating the latter events in the energy markets. After employing a variety of unit root and cointegration tests, the long-run relationship is examined via the autoregressive distributed lag (ARDL) cointegration technique, along with the Toda-Yamamoto (1995) causality test. Our results indicate that there is a long-run relationship with a unidirectional causality running from the Dubai crude oil market to the US natural gas market. A variety of post specification tests indicate that the selected ARDL model is well-specified, and the results of the Toda-Yamamoto approach via impulse response functions, forecast error variance decompositions, and historical decompositions with generalized weights, show that the Dubai crude oil price retains a positive relationship and affects the US natural gas price.

## 1. Introduction

The linkages between crude oil and natural gas prices have remained a subject of ongoing research whether information transition runs form crude oil to natural gas. Serletis and Herbert (1999) worked on the North American prices, Villar and Joutz (2006) on the Henry Hub and Western Texas Intermediate (WTI), Panagiotidis and Rutledge (2007) on the relationship between UK wholesale gas prices and the Brent oil price, Hartley et al. (2008) on Henry Hub, fuel oil, and WTI, Nick and Thoenes (2014) on the connection of the Brent oil price with Germany natural gas price, Atil (2014) on WTI, gasoline, and Henry Hub natural gas, Brigida (2014) on US natural gas and oil prices, and Gatfaoui (2016) the joint link between U.S. oil and gas markets others. Recently, Lin and Li (2015) investigated both price and volatility spillover effects for Crude oil and natural gas markets of US, Europe, and Japan, for regional segmentation and different pricing mechanisms of natural gas. Batten et al. (2017) examined daily settlement prices of the nearby contract to the spot date of Henry Hub natural gas and crude oil futures traded on the New York Mercantile Exchange (NYMEX). Although most of the works are focused on the major benchmark markets several works have started to concentrate on the linkages of other crude oil markets.

Towards the Dubai crude oil literature Lu et al. (2008) studied the WTI futures and spot prices, Brent futures and spot prices, Dubai, Tapis, and Minas oil spot prices, Hammoudeh et al. (2008) examined the dynamic relationships between pairs of four oil benchmark prices (WTI, Brent, Dubai, and Maya), Kang et al. (2009) examined the efficiency in the out-of-sample volatility forecasts of a variety of GARCH volatility models for three crude oil markets Brent, Dubai, and WTI, Reboredo (2011) examined the WTI, Brent, Dubai, and Maya spot market prices, thus supporting the hypothesis that the oil market is 'one great pool', in contrast with the hypothesis stating that the oil market is regionalized. Jin et al. (2012) examined crude oil markets integration on the second moment for WTI, Dubai and Brent futures contracts, Lu et al. (2014) applied time-varying Granger causality tests to examine time-varying information spillover among global crude oil markets. Ji and Fan (2015) explored the leading/lagging relationship between the world's major crude oils, WTI, Brent, Dubai, Tapis, and Nigeria. Ding et al. (2016) considered the causal relationships between WTI and Dubai crude oil returns and five stock index returns (S&P 500, Nikkei, Hang Seng, Shanghai, and KOSPI) within the quantile causality framework. Kuck and Schweikert (2017) examined the long-run equilibrium relationships between major crude oil prices, WTI, Brent, Bonny Light, Dubai, and Tapis, via a Markov-switching vector error correction model to capture changing roles of crudes in the world crude oil market, and a changing degree of market integration. They also concluded that the crude oil market is globalized, with Dubai to be the only weakly exogenous price in all regimes, indicating its significant role as a benchmark price.

In this work, we examine the interrelationship between the Dubai crude oil and the US natural gas. Applying a broad variety of unit root tests, taking into account that most tests are sensitive to several types of non-stationarity, the autoregressive distributed lags (ARDL) specification along with the bound test is identified as the appropriate specification. The results indicate cointegration running from the Dubai crude oil to the US natural gas, and not vice-versa, and application of the Toda-Yamamoto procedure along with the impulse response functions identify a unidirectional causality from the Dubai crude oil to the US natural gas. Several posterior tests show that the model is well specified.

The remainder of the paper is organized as follows. Section 2 describes the data used in this work with a preliminary analysis of stylized facts and descriptive statistics, and the econometric methodology followed is presented. Section 3 presents the empirical results along with the relevant discussion, and Section 4 offers the concluding remarks.

## 2. Data and econometric methodology

### 2.1. The data

The dataset was obtained from the International Monetary Fund (IMF) commodity data from 1992M01 to 2018M10. The price of the Dubai crude oil is given in nominal US dollars per bbl, and the price of US natural gas is given in nominal US dollars per mmbtu. Due to the fact that the intitial series possess large heteroskedasticity a practical approach is to work with the natural logarithm of the series. Fig.1 shows the time series under consideration, the price of the Dubai crude oil (dubai, top left), the price of the US natural gas (gasus, bottom left), the natural log of the price of the Dubai crude oil (lndubai, top right), and the natural logarithm of the price of the US natural gas (lngasus, bottom right).

Table 1 shows the descriptive statistics and stylized facts for the series returns, defined as the difference of the natural logarithms of the initial series. We report on the first four moments that is the mean, the standard deviation, the skewness and the kurtosis, the Jarque-Bera test for normality (JB), the ARCH effect, and the Ljung-Box statistics on both the returns (LB) and the square returns (LB-2) for 10 lags. All returns possess autocorrelation, heteroskedasticity, and deviation from normality.

After the Global Financial Crisis (GFC) the rapidly increasing demand for energy by emerging markets including China and India, along with the production decrease by the Organization of Petroleum Exporting Countries (OPEC) in Middle East, resulted in high oil prices for three years. The low-interest rates during that time, gave banks and private equity investors the opportunity to lend to shale oil and gas companies. However, unlike oil, natural gas is a local market because it is difficult to transport it without a pipeline, unless it is liquefied which is costly. Therefore, the amount produced in the US and Canada causes a direct effect on the local natural gas price, and this is seen in Fig. 1 (bottom, left) where a decoupling effect of the price of US natural gas to the price of Dubai oil is observed after the GFC[1].

## 2.2. The methodology

### 2.2.1. ARDL Cointegration Analysis and Bounds Testing Approach

The autoregressive distributed lag (ARDL) model, introduced originally by Pesaran and Shin (1999) and further extended by Pesaran et al. (2001), takes into account a single cointegrating equation. In contrast to the usual cointegration framework of Granger (1981), Engle-Granger (1987), and Johansen (1988, 1991) the ARDL approach has the advantage that it does not require all variables to be I(1) and it is still applicable if there are I(0) and I(1) variables in the dataset. The augmented general ARDL$(p, q_1, q_2, \ldots, q_k)$ model specification is given as follows,

$$a(L,p)y_t = a_0 + \sum_{i=1}^{k} \beta_i(L, q_i)x_{it} + \lambda' w_t + u_t, \quad \forall t = 1, 2, \ldots, n \tag{1}$$

$$a(L,p) = 1 - a_1 L - a_2 L^2 - \cdots - a_p L^p \tag{2}$$

---

[1] Extending the analysis on the interrelationship of the Dubai Fateh oil with the rest two natural gas benchmarks that is, the Russian natural gas border price in Germany, and the Indonesian Liquefied Natural Gas price in Japan, canonical cointegration is observed. There is a unidirectional long-run causality running from the Dubai Fateh Oil to the natural gas markets, differentiating them for the US natural gas due to the locality of the market. Due to space limitations, the results are available upon request.

$$\beta_i(L, q_i) = \beta_{i0} - \beta_{i1}L - \beta_{i2}L^2 - \cdots - \beta_{iq_i}L^{q_i}, \quad \forall i = 1, 2, \ldots, k \tag{3}$$

where $y_t$ is the dependent variable, $a_0$ is the constant term, $L$ is the lag operator, $w_t$ is a vector of deterministic variables that might include an intercept term, time trend, seasonal dummies, or exogenous variables with fixed lags, $x_t$ is the $k$-dimensional forcing variables, which should not be cointegrated among themselves, and $\varepsilon_t \sim iid(0, \sigma^2)$ is the vector of stochastic error terms with zero mean and constant variance-covariance.

The long run coefficients are estimated by,

$$\pi = \frac{\hat{\lambda}(\hat{p}, \hat{q}_1, \hat{q}_2, \ldots, \hat{q}_k)}{1 - \hat{a}_1 - \hat{a}_2 - \cdots - \hat{a}_{\hat{p}}} \tag{4}$$

where $\hat{\lambda}(\hat{p}, \hat{q}_1, \hat{q}_2, \ldots, \hat{q}_k)$ denotes the OLS estimates of $\lambda'$ for the selected ARDL model. After that, the error correction model related to the specified ARDL$(p, q_1, q_2, \ldots, q_k)$ can be obtained using the lagged levels and the first difference of $y_t, x_{1t}, x_{2t}, \ldots, x_{kt}$, and $w_t$,

$$\Delta y_t = -a(1, \hat{p})ECM_{t-1} + \sum_{i=1}^{k} \beta_{i0} x_{it} + \lambda' \Delta w_t - \sum_{j=1}^{\hat{p}-1} a_j^* \Delta y_{t-1} - \sum_{i=1}^{k} \sum_{j=1}^{\hat{q}_j-1} \beta_{ij} \Delta x_{i,t-1} + \varepsilon_t \tag{5}$$

where $ECM$ is the error correction term defined as $ECM_t = y_t - \hat{a} - \sum \hat{\beta}_{i0} x_{it} - \lambda' w_t$.

After that, the hypothesis that the coefficients of the lag level variables are zero is tested, under the null that the long run relationship does not exist. The critical values of the F-statistics for different number of variables, and whether the ARDL model contains an intercept and/or trend are available in Pesaran et al. (2001). They give two sets of critical values, one set assuming all the variables are I(0) (lower critical bound of no cointegration among the underlying variables), and another assuming all the variables in the ARDL model are I(1) (upper critical bound) indicating cointegration among the underlying variables). However, according to Narayan (2005) the existing critical values in Pesaran et al. (2001) cannot be applied for small sample sizes, as they are based on large sample sizes, and a set of critical values for small sample sizes ranging from 30 to 80 observations is provided.

### 2.2.2. Toda –Yamamoto Causality Test

The dynamic Granger causality can be captured from the vector error correction model derived from the long-run cointegrating relationship Granger (1988). The Granger causality proposed by Granger (1969) has probable shortcomings of specification bias and spurious regression. Engle and Granger (1987) have defined two variables to be cointegrated if a linear combination of them is stationary but each variable is not stationary. Engel and Granger (1987) pointed out that while these two variables are non-stationary and cointegrated, the standard Granger causal inference might be invalid.

To mitigate these problems Toda and Yamamoto (1995), and Dolado and Lutkepohl (1996) (TYDL) based on augmented vector autoregressive (VAR) modeling, introduced a modified Wald test statistic. This procedure has been found to be superior to ordinary Granger causality tests since it does not require pre-testing for the cointegrating properties of the system and thus avoids the potential bias associated with unit roots and cointegration tests. It can be applied regardless of whether a series is I(0), I(1) or I(2), non-cointegrated, or cointegrated of an arbitrary order. The TYDL approach requires finding the maximum order of integration $d_{max}$ of the series that are to be incorporated in the model, applying conventional unit root tests, where the number of optimal lags is determined by selection criteria like AIC, BIC, or SIC. The VAR has to be well specified in terms of the AR unit root graph, the VAR residual serial correlation LM-statistics, and the VAR residual normality tests. The TYDL approach over-fits the underlying model with the additional maximum order of integration $d_{max}$ treated as an exogenous variable,

$$y_t = \mu_0 + \sum_{i=1}^{k} a_{1i} y_{t-i} + \sum_{i=k+1}^{d_{max}} a_{2i} y_{t-i} + \sum_{i=1}^{k} \beta_{1i} x_{t-i} + \sum_{i=k+1}^{d_{max}} \beta_{2i} x_{t-i} + \varepsilon_{1t} \qquad (6)$$

$$x_t = \varphi_0 + \sum_{i=1}^{k} \gamma_{1i} x_{t-i} + \sum_{i=k+1}^{d_{max}} \gamma_{2i} x_{t-i} + \sum_{i=1}^{k} \delta_{1i} y_{t-i} + \sum_{i=k+1}^{d_{max}} \delta_{2i} y_{t-i} + \varepsilon_{2t} \qquad (7)$$

The modified Wald test follows asymptotically a Chi-square distribution where the degrees of freedom are equal to the number of lags $k + dmax$. Rejection of the null hypothesis entails the rejection of Granger causality.

## 3. Results and discussion

### 3.1. Stationarity properties and order of integration

The most important issue when dealing with the cointegration of time series is whether or not both series are non-stationary I(1), both series are stationary I(0), or each of the series possesses a different order of integration. A non-stationary stochastic process might be trend stationary process (TSP) or difference stationary process (DSP). If it's not clear whether the process is TSP and is treated as DSP the result is over-differentiation, while if the series is DSP and treated as TSP the result is under-differentiation. However, when dealing with time-series there might be an overlapping of a variety of non-stationarities that might include unit-roots, structural breaks, level shifts, seasonal cycles or seasonality in the sixth moment, or a changing variance. These effects make the usual unit-root tests incapable of properly identifying the data generation process and more advanced tests are in order. Although ARDL cointegration technique does not require pre-testing for unit roots, to avoid ARDL model crash in the presence of an integrated stochastic trend of I(2), unit root test should be carried out to examine the number of unit roots in the series under consideration.

Table 2 shows a variety of unit root tests in order to identify the integration order of the series under consideration. We report on the augmented Dickey-Fuller (ADF), the Dickey-Fuller using generalized least squares rationale (DFGLS), the Phillips-Perron (PP), the Kwiatkowski–Phillips–Schmidt–Shin (KPSS), and the Elliot, Rothenberg, and Stock Point Optimal (ERS). Incorporating structural breaks in the unit-root tests we also report on the Vogelsang and Perron (1998), and Kim and Perron (2009) test with an innovation outlier including a trend and an intercept (P-IO), the Perron test with an additive outlier including a trend and an intercept (P-AO), the Zivot-Andrews (1992) test with one structural break including an intercept and a trend, and the Lee-Strazicich (2003) with two structural breaks, crash with level-shift only (LSC), and the most general break model (LSB). In all test except the KPSS the null hypothesis is that the series possess a unit root.

The tests indicate the presence of non-stationarity for the lndubai variable classifying it as I(1) series. Most of the tests indicate the presence of non-stationarity for the lngasus variable, however; the Lee-Strazicich break test with 2 breaks rejects marginally at the 1% level the null hypothesis of a unit-root. Therefore, the lngasus variable can be classified as a "near unit root" series.

**3.2. Cointegration tests**

Table 3 shows the VAR lag order selection criteria for the log-levels of the variables for 12 lags with exogenous variables a constant and a trend. The parsimonious criteria Schwarz (SC) and Hannan-Quinn (HQ) indicate 2 lags, the Akaike (AIC) and the final prediction error (FPE) 3 lags, and the likelihood ratio (LR) 11 lags. Table 4 shows a variety of cointegration tests, Panel A reports on the Engle-Granger (1987) test, and Panel B on the Phillips-Ouliaris (1990) test, which are both residual-based tests. Panel C reports on the Johansen trace and maximum eigenvalue tests, and Panel D on the Gregory and Hansen (1996a, 1996b) residual cointegration test. In all tests the null hypothesis is no cointegration. The Engle-Granger test rejects marginally the null of no cointegration for both cases as a dependent variable, while in the same sense the Phillips-Ouliaris test accepts the null hypothesis of no cointegration when the lndubai variable is the dependent variable, but rejects the null hypothesis for the lngasus variable. The Johansen methodology in both forms trace and maximum eigenvalue, accepts the null hypothesis that there is no cointegrating vector, however; the information criteria by rank and model identify a possible cointegration case via the AIC criterion. The Gregory-Hansen test analyses the cointegration between variables with structural breaks (switching regime), estimating a single equation cointegrating regression by least squares allowing for structural change of unknown time. When lngasus is the dependent variable the null hypothesis of no cointegration is rejected in all cases, however; the null hypothesis is not rejected when lngasus is the dependent variable for the most general cases of break regime and break regime with trend.

**3.3. ARDL Cointegration Analysis and Bounds Testing Approach**

Tables 5-6 show the results of two representations for the ARDL(1,7) specification working down from a maximum 8 lags for the dependent variable lngasus and the regressor lndubai, with the Akaike criterion as an automatic selection. Table 5 shows the result of the intertemporal dynamics regression of the ARDL(1,7) specification, and Table 6 shows the results for the ARDL error correction regression, where the error correction term is,

$$EC = LNGAUSUS - 0.3326[10.476] \times LNDUBAI \tag{8}$$

The bounds test rejects the null hypothesis of no-levels relationship with an F-bounds test statistic of 5.0529, and a t-bounds test statistics of -3.1652. The assorted 5% confidence levels are 3.15 I(0) and 4.11 I(1) for the F-bounds test, and -1.95 I(0) and -2.6 I(1) for the t-bounds test. The error correction coefficient (Table 6) is negative and statistically significant indicating a slow adjustment of 5.67% per month. Performing a Wald test whether the lagged regressors of the lndubai are zero, the null hypothesis is rejected with F-statistic=4.1894[0.0011] identifying a short run Granger causality from lndubai to lngasus.

Alternating the variables and using lndubai as the dependent variable and lngasus as the regressor, in an ARDL(2,1) specification the bound test accepts the null hypothesis of no levels relationship with an F-statistic of 0.4036, and a t-bounds test statistic of -0.546774.

Examining whether the model is well specified, there is no serial correlation up to 6 lags using the Breusch-Godfrey Serial Correlation LM Test, resulting in an F-statistic=0.9223[0.4791]. The Ramsey RESET test, using the squares of the fitted values as omitted variables to test for model misspecification via non-linear combinations of the explanatory variables, accepts the null hypothesis with an F-statistic=0.0586[0.8088]. The Breusch-Pagan-Godfrey and White heteroskedasticity tests reject the null hypothesis of homoscedasticity at the 5% level but not at the 1% level, with F-statistic=2.462495[0.0101], and F-statistic=2.288207[0.0170] respectively. However, the rest of the tests, Harvey's F-statistic=1.1170[0.3502], Glejser's F-statistic=1.7874[0.070], and the ARCH effect with 10 lags F-statistic=1.420270[0.1704] accept the null hypothesis of no heteroskedasticity.

The existence of a stable and predictable relationship between the lngasus and lndubai variables is a necessary condition when it comes to policy practices. The stability of the long-run coefficients is used to form the error-correction term in conjunction with the short run dynamics. In view of this we apply the CUSUM test involving the calculation of a cumulative sum of the recursive residuals, and the CUSUMQ test on the squares of the recursive residuals (Fig. 2). The plots of CUSUM and CUCUMQ statistics stay within the 5% significance level indicating that the estimated coefficients are stable.

### 3.4. Toda –Yamamoto Causality Test

The Toda-Yamamoto causality test is realized by building a VAR model with 3 endogenous lags (selecting the AIC and FPE choice from Table 3) and one exogenous lag since the maximum order of integration is $dmax = 1$,

$$lndubai_t = \sum_{i=1}^{3} a_i lndubai_{t-i} + a_4 lndubai_{t-4} + \sum_{i=1}^{3} \beta_i lngasus_{t-i} + \beta_4 lngasus_{t-4} + \varepsilon_{1t} \quad (9)$$

$$lngasus_t = \sum_{i=1}^{3} \gamma_i lndubai_{t-i} + \gamma_4 lndubai_{t-4} + \sum_{i=1}^{3} \delta_i lngasus_{t-i} + \delta_4 lngasus_{t-4} + \varepsilon_{2t} \quad (10)$$

Table 7 shows the VAR Granger Causality/Block Exogeneity Wald tests where a unidirectional causality from lndubai to lngasus is identified but not vice-versa. The Breusch-Godfrey serial Correlation LM test accepts the null hypothesis of no serial correlation up to 7 lags with F-statistic=1.8132[0.0843], and the VAR Residual Portmanteau test shows no significant autocorrelation up to 7 lags. The estimated VAR is stable (stationary) if all inverse roots of the characteristic polynomial have modulus less than one and lie inside the complex unit circle (Fig. 3). VAR models represent the correlations among a set of variables, and they are often used to analyze certain aspects of the relationships between the variables of interest.

The response of a variable to a unit shock in another variable, depicted graphically, can give a visual impression of the dynamic interrelationships within the system. Fig. 4 shows the Impulse Response Functions (IRF) for the system of variables under consideration. The concern about ordering in a VAR is especially due to the impulse response (IRF) specification. Since a Cholesky decomposition is used to realize the IRFs, the order of the variables has to been taken care of. In case of a lower triangular matrix the variables are ordered via an exogeneity criterion, so the lndubai variable is put first and the lngasus variable second. The IRFs show that a positive shock at the lndubai price has a positive impact and affects the lngasus price (Fig. 4, bottom) but the opposite does not hold (Fig. 4 top).

The Forecast Error Variance Decomposition (FEVD) shows how much of the future uncertainty of one time series is due to future shocks from other time series in the system. This evolves over time, so the shocks on time series may not be very important in the short-run but very important in the long run. While IRFs trace the effects of a shock to one endogenous variable on to the other variables in a VAR system, FEVD distinguishes the variation in an endogenous variable into the component shocks to the VAR, providing information about the relative importance of each random innovation in affecting the variables in the VAR system. Fig. 5 shows the FEVD graphs for the lndubai-lngasus system. It is shown that on the long run lndubai can account for about 18% of the future uncertainty of the lngasus variable (Fig. 5 top, right) but the opposite does not hold (Fig. 5 top, left).

Another widely reported output from VAR models is the historical decomposition. The idea is that all series included in the VAR can be fully decomposed into the contribution of different shocks and an exogenous component, referred to as the baseline projection. If the contributions of all the shocks at any time together with the baseline projection are summed up, we recover the original time series. In the historical decomposition one investigates how differently would variables have evolved if particular histories of shocks have instead occurred. While there are some similarities between this and the decomposition of variance, there are also some key differences, where the most important is that this is not a partitioning of any quantity. Fig. 5 shows the historical decompositions using generalized impulses indicating that there is a borderline contribution to the lndubai baseline from lngasus (Fig. 5 bottom left), while is not the same for the contribution of the lndubai variable to the baseline of lngasus (Fig. 5 bottom right).

## 4. Conclusion

Crude oil and natural gas play strategic roles in socio-economic development around the world, and global demand for energy is continuously rising because developed countries consume large amounts of energy, while demand in developing countries is increasing. They are the major energy sources for the global economy, and understanding their price dependencies is important for a variety of reasons including energy investment, policy decisions, portfolio diversification and hedging capabilities. Governments tighten up environmental regulations, seeking alternative energy sources to meet energy demand via reduction of the dependency on oil, with natural gas representing an economically viable alternative solution. Under these conditions, the relationship between crude oil and natural gas prices also reveals the degree of substitutability and the level of competition among energy producers.

In this work we have examined the interrelationship between the Dubai crude oil and the US natural gas, using an updated dataset incorporating the latter events in the energy markets. A variety of unit root and cointegration tests identify the autoregressive distributed lag (ARDL) cointegration technique along with the bound test as the proper specification to work with. Our results indicate that there is a long-run relationship with a unidirectional causality running from the Dubai crude oil market to the US natural gas market. A variety of post specification tests indicate that the selected ARDL model is well-specified, and the results of the Granger causality test, following the Toda-Yamamoto approach and impulse response functions (IRF), show that the Dubai crude oil price retains a positive relationship and affects the US natural gas price.

**Figure 1** The time series under consideration (left, top and bottom) and the natural logarithms of the series (right, top and bottom)

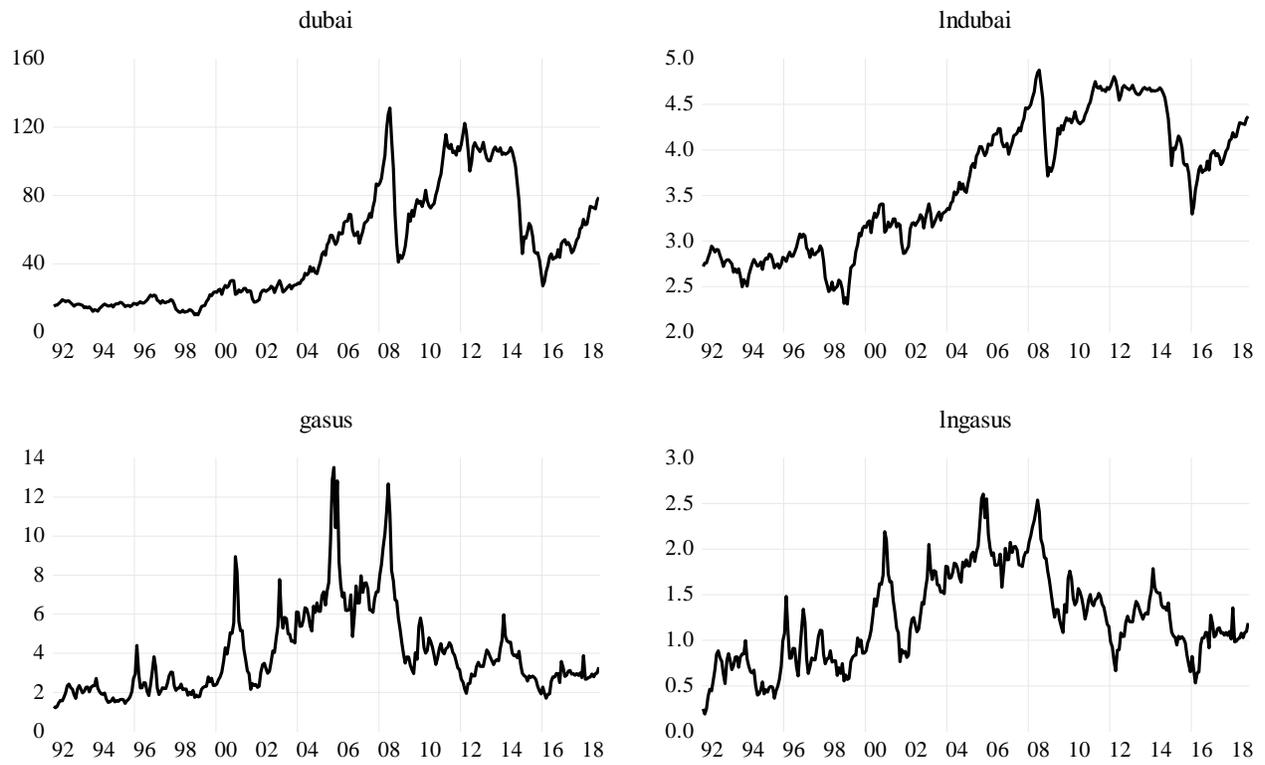

**Figure 2** Cumulative sums of the recursive residuals (top) and of the squares of the recursive residuals (bottom)

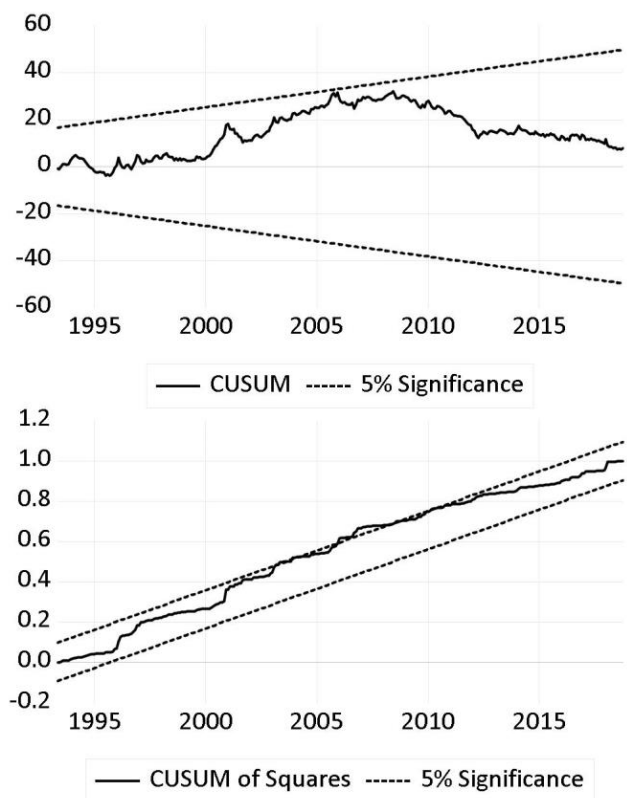

**Figure 3** Inverse Roots of the AR characteristic polynomial

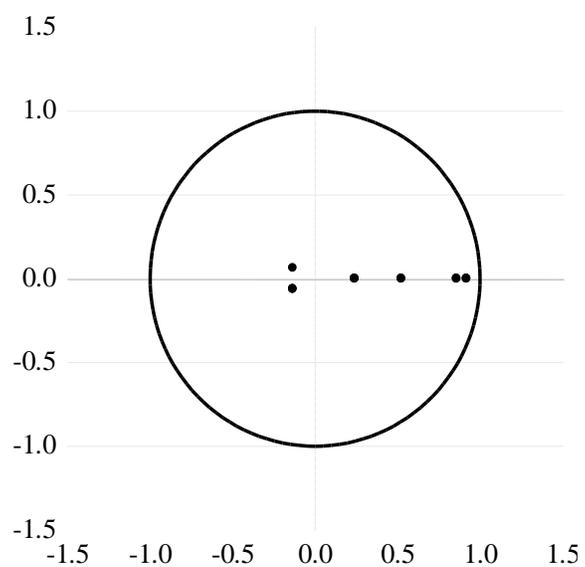

**Figure 4** Impulse Responses of lndubai to lngasus (top) and lngasus to lndubai (bottom) to Cholesky one S.D. (d.f. adjusted) Innovations ± 2 S.E.

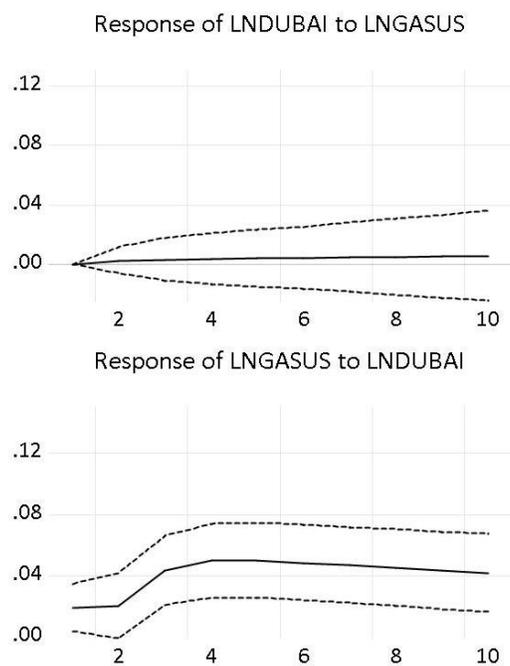

**Figure 5** Forecast error variance decomposition (top) and historical decomposition via generalized impulses (bottom)

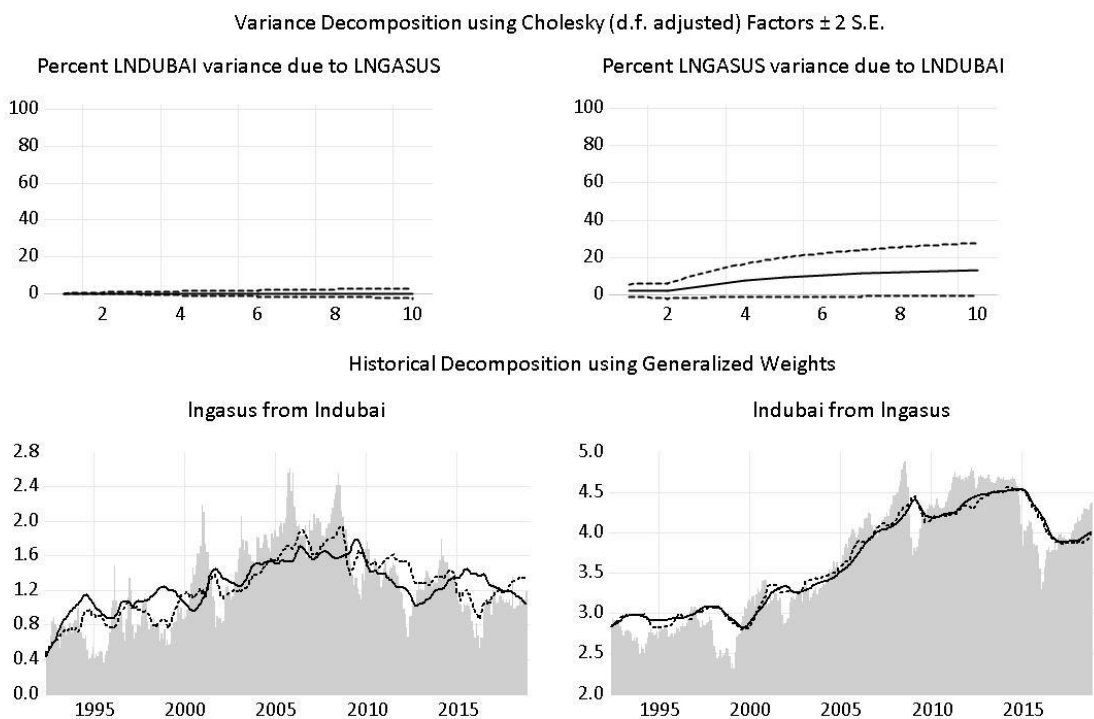

**Table 1** Descriptive statistics and stylized facts

| returns | mean | stdev | skewness | kurtosis | JB | ARCH(10) | LB(10) | LB-2(10) |
|---|---|---|---|---|---|---|---|---|
| dlndubai | 0.0051 | 0.0813 | -0.9425* | 5.2470* | 115.06* | 8.8135* | 43.638* | 84.835* |
| dlngasus | 0.0029 | 0.1346 | 0.08713 | 4.1522* | 18.164* | 2.0392* | 19.560* | 23.135* |

*Notes: (\*) denotes statistical significance at the 5% critical level. JB is the statistic for the null of normality; LB(10) denotes the Ljung–Box test statistic for the residuals, LB-2(10) denotes the Ljung–Box test statistic for the squared residuals, and ARCH(10) the test for residual heteroskedasticity, for 10 lags.*

**Table 2** Unit root tests on the log-series and the returns

| Panel A: (log-levels) | ADF | DFGLS | PP | KPSS | ERS | P-IO | P-AO | ZA | LS-C | LS-B |
|---|---|---|---|---|---|---|---|---|---|---|
| lndubai | -2.42 | -2.40 | -2.02 | 0.26* | 7.83 | -4.20 | -4.18 | -4.18 | -2.94 | -5.01 |
| lngasus | -2.74 | -2.06 | -2.80 | 0.44* | 11.2 | -5.12 | -4.36 | -5.14 | -3.39 | -6.07* |
| Panel B: (returns) | ADF | DFGLS | P.P. | KPSS | ERS | PIO | PAO | ZA | LSC | LSB |
| dlndubai | -13.4* | -13.2* | -13.2* | 0.06 | 0.64* | -14.0* | -14.1* | -13.7* | -12.6* | -13.2* |
| dlngasus | -16.9* | -16.1* | -16.9* | 0.02 | 0.61* | -17.3* | -17.4* | -17.0* | -8.02* | -11.4* |

*Notes: (\*) denotes statistical significance at the 5% critical level. ADF is the Augmented Dickey Fuller test, DFGLS the Dickey-Fullerwith GLS test, ERS is the Elliot, Rothenberg, and Stock Point Optimal test, P.P. is the Phillips-Perron test, and KPSS is the Kwiatkowski, Phillips, Schmidt, and Shin unit root test. P-IO is the Perron innovation outlier test with one break, P-AO is the Perron additive outlier test with one break, ZA is the Zivot-Andrews test with one break, LS-C is the Lee-Strazicich crash test with two breaks with level-shift only, and LS-B is the Lee-Strazicich break test with two breaks.*

**Table 3** VAR Lag Order Selection Criteria

| Lag | LogL | LR | FPE | AIC | SC | HQ |
| --- | --- | --- | --- | --- | --- | --- |
| 0 | -305.8749 | NA | 0.025310 | 1.999193 | 2.047407 | 2.018467 |
| 1 | 523.3988 | 1637.147 | 0.000123 | -3.325154 | -3.228726 | -3.286606 |
| 2 | 538.0935 | 28.82068 | 0.000115 | -3.394152 | -3.249510* | -3.336330* |
| 3 | 542.6833 | 8.942696 | 0.000115* | -3.397957* | -3.205102 | -3.320862 |
| 4 | 543.7194 | 2.005292 | 0.000117 | -3.378835 | -3.137766 | -3.282466 |
| 5 | 545.1841 | 2.815978 | 0.000119 | -3.362478 | -3.073195 | -3.246835 |
| 6 | 545.6733 | 0.934258 | 0.000122 | -3.339828 | -3.002331 | -3.204911 |
| 7 | 550.9704 | 10.04732 | 0.000121 | -3.348196 | -2.962485 | -3.194005 |
| 8 | 553.9490 | 5.611370 | 0.000121 | -3.341607 | -2.907682 | -3.168142 |
| 9 | 555.0339 | 2.029742 | 0.000124 | -3.322799 | -2.840661 | -3.130061 |
| 10 | 561.8204 | 12.60987 | 0.000121 | -3.340777 | -2.810425 | -3.128765 |
| 11 | 567.4917 | 10.46447* | 0.000120 | -3.351560 | -2.772994 | -3.120273 |
| 12 | 570.5629 | 5.627226 | 0.000121 | -3.345567 | -2.718787 | -3.095007 |

*Notes: (\*) denotes statistical significance at the 5% critical level. LogL is the log-likelihood, LR is the likelihood ratio, FPE is the final prediction error and AIC, SC, and HQ are the Akaike, Schwarz and Hannan-Quinn information criteria.*

**Table 4** Cointegration tests

Panel A: Engle-Granger cointegration test

| Dependent | t-statistic | Prob.* | z-statistic | Prob.* |
|---|---|---|---|---|
| LNDUBAI | -2.9371 | 0.0330 | -15.752 | 0.0481 |
| LNGASUS | -3.0788 | 0.0227 | -17.328 | 0.0335 |

Panel B: Phillips-Ouliaris cointegration test

| Dependent | t-statistic | Prob.* | z-statistic | Prob.* |
|---|---|---|---|---|
| LNDUBAI | -2.9055 | 0.0358 | -15.326 | 0.0531 |
| LNGASUS | -3.0692 | 0.0233 | -17.149 | 0.0349 |

Panel C: Johansen cointegration test

Unrestricted Cointegration Rank Test (Trace)

| Hypothesized No. of CE(s) | Eigenvalue | Trace Statistic | 0.05 Critical Value | Prob.* |
|---|---|---|---|---|
| None | 0.0328 | 10.989 | 12.320 | 0.0827 |
| At most 1 | 0.0009 | 0.2893 | 4.1299 | 0.6517 |

Unrestricted Cointegration Rank Test (Maximum Eigenvalue)

| Hypothesized No. of CE(s) | Eigenvalue | Max-Eigen Statistic | 0.05 Critical Value | Prob.* |
|---|---|---|---|---|
| None | 0.0328 | 10.699 | 11.224 | 0.0618 |
| At most 1 | 0.0009 | 0.2893 | 4.1299 | 0.6517 |

Panel D: Gregory-Hansen cointegration test

| Dependent: lndubai | ADF | Zt | Za |
|---|---|---|---|
| Break Level | -5.20* | -5.34* | -44.42* |
| Break Regime | -4.97* | -5.08* | -40.70 |
| Break Regime with Trend | -4.43 | -4.80 | -31.72 |

| Dependent: lngasus | ADF | Zt | Za |
|---|---|---|---|
| Break Level | -5.61* | -5.65* | -50.38* |
| Break Regime | -5.77* | -5.91* | -54.41* |
| Break Regime with Trend | -6.05* | -6.27* | -63.94* |

*Notes: (\*) denotes MacKinnon (1996) p-values, and for the Gregory-Hansen cointegration test statistical significance at the 5% critical level.*

**Table 5** Intertemporal dynamics regression results of the ARDL(1,7) specification

Dependent Variable: LNGASUS

| Variable | Coefficient | Std. Error* | t-Statistic | Prob. |
|---|---|---|---|---|
| LNGASUS(-1) | 0.9432 | 0.0183 | 51.284 | 0.0000 |
| LNDUBAI | 0.2802 | 0.1379 | 2.0311 | 0.0431 |
| LNDUBAI(-1) | -0.3194 | 0.1867 | -1.7105 | 0.0882 |
| LNDUBAI(-2) | 0.3189 | 0.1543 | 2.0663 | 0.0396 |
| LNDUBAI(-3) | -0.1899 | 0.1599 | -1.1877 | 0.2359 |
| LNDUBAI(-4) | 0.0345 | 0.1362 | 0.2537 | 0.7999 |
| LNDUBAI(-5) | -0.2550 | 0.1413 | -1.8042 | 0.0722 |
| LNDUBAI(-6) | 0.3815 | 0.1473 | 2.5886 | 0.0101 |
| LNDUBAI(-7) | -0.2319 | 0.0848 | -2.7352 | 0.0066 |

*Notes: (*) denotes HAC standard errors and covariance.*

**Table 6** ARDL(1,7) error correction regression representation

Dependent Variable: D(LNGASUS)

| Variable | Coefficient | Std. Error* | t-Statistic | Prob. |
|---|---|---|---|---|
| D(LNDUBAI) | 0.2802 | 0.0936 | 2.9919 | 0.0030 |
| D(LNDUBAI(-1)) | -0.0580 | 0.0966 | -0.6008 | 0.5484 |
| D(LNDUBAI(-2)) | 0.2608 | 0.0966 | 2.7005 | 0.0073 |
| D(LNDUBAI(-3)) | 0.0709 | 0.0966 | 0.7339 | 0.4636 |
| D(LNDUBAI(-4)) | 0.1054 | 0.0966 | 1.0913 | 0.2760 |
| D(LNDUBAI(-5)) | -0.1495 | 0.0967 | -1.5457 | 0.1232 |
| D(LNDUBAI(-6)) | 0.2319 | 0.0938 | 2.4729 | 0.0139 |
| CointEq(-1)* | -0.0567 | 0.0178 | -3.1841 | 0.0016 |

*Notes: (*) denotes HAC standard errors and covariance.*

**Table 7** VAR Granger Causality/Block Exogeneity Wald Tests

Dependent variable: LNDUBAI

| Excluded | Chi-sq. | df | Prob. |
|---|---|---|---|
| LNGASUS | 0.5428 | 3 | 0.9094 |
| All | 0.5428 | 3 | 0.9094 |

Dependent variable: LNGASUS

| Excluded | Chi-sq | df | Prob. |
|---|---|---|---|
| LNDUBAI | 13.530 | 3 | 0.0036* |
| All | 13.530 | 3 | 0.0036* |

*Notes: (*) denotes statistical significance at the 5% critical level.*